\documentclass[aps,prl,twocolumn,groupedaddress]{revtex4}

\usepackage{graphicx}
\usepackage{amssymb}
\usepackage{epstopdf}
\def\be{\begin{equation}}   \def\ee{\end{equation}}
\def\eq#1{{Eq.(\ref{#1})}}    \def\fig#1{{Fig.\ref{#1}}}

\def\F{{\cal F}}           \def\A{{\cal A}}      \def\N{{\cal N}}   
   
\def\fsk{{f_{\sigma,\kappa}}}   \def\fsk{{f_{\sigma,\kappa}}}

\def\bgamma{{\bar\gamma}}     \def\bsigma{{\bar\sigma}}      \def\bkappa{{\bar\kappa}}         

\def\bks{{\bar{K_s}}}                 

    \def\ab{{\alpha_{bud}}}      \def\bb{{\beta_{bud}}}

\begin{document}

\title{Budded membrane microdomains as regulators for cellular tension}
\author{Pierre Sens$^{1,2}$ and Matthew S. Turner$^3$}
\affiliation{$^1$ UMR 168 - Physico-Chimie Curie - Institut Curie, $^2$ UMR 7083 - Physico-Chimie Th\'eorique - ESPCI\\
10 rue Vauquelin, 75231 Paris Cedex 05 - France\\
pierre.sens@espci.fr\\
$^3$ Department of Physics, University of Warwick, Coventry, CV47AL - UK }

\date{\today}

%\pacs{
%87.16.-b Subcellular structure and processes
%87.16.Dg Membranes, bilayers, and vesicles
%67.40.Fd Dynamics of relaxation phenomena
%}

%\keywords{}

\begin{abstract}
We propose a mechanism for the control of the mechanical properties of the membrane of living cells that is based on the exchange of membrane area between the cell membrane and a membrane reservoir.  The reservoir is composed of invaginated membrane microdomains which are liable to flatten upon increase of membrane strain, effectively controlling membrane tension. We show that the domain shape transition is first order, allowing for coexistence between flat and invaginated domains. During coexistence, the membrane tension is controlled by the domains elasticity and by the kinetics of the shape transition. We show that the tension of the plasma membrane of living cells is inherently transient and dynamical, and that valuable insights into the organization of the cell membrane can be obtained by studying the variation of the cell membrane tension upon mechanical perturbation. 
\end{abstract}

\maketitle

Endocytosis, exocytosis, cell motility and many other crucial cellular processes are known to be influenced by the tension of the cell membrane\cite{sheetz}. The force needed to pull a tubular membrane tether from the cell with an optical trap is often used as a probe of the mechanical state of the membrane\cite{sheetz_cytoskel}. The mechanical response of living cells to such perturbation involves cytoskeleton deformation, the breaking of membrane-cytoskeleton bonds\cite{sheetz_pip2} and changes in membrane morphology. The level of membrane tension in cells is thought to primarily reflects cytoskeleton anchoring to the membrane\cite{sheetz_cytoskel}. Nevertheless, there is experimental evidence that Surface Area Regulation (SAR) occurs \cite{morris_SAR} and is able to buffer the increase of tension upon mechanical perturbation. SAR could be achieved by the transfer of membrane area from a reservoir to the plasma membrane, during which the level of membrane tension is at least partly controlled by the dynamical response of  the reservoir. One possible manifestation of the exchange of membrane with a reservoir is the presence of a plateau in the force-extension curve of tether extraction experiments, as observed in \cite{sheetz_buffer}. The plateau ends when the putative membrane reservoir is emptied. In this particular experiment the reservoir is able to react quite fast to the perturbation ($<0.1$s), which could indicate {\em i}) that the reservoir is permanently connected to the plasma membrane and {\em ii}) that the regulation is purely physical and does not involve biological signaling. 

Our goal in this paper is to study the influence of membrane morphology, and in particular the flattening of invaginated "bud-like" domains, on the mechanical response of the membrane. Such domains include {\em caveolae} that form $\Omega$ shaped invaginations\cite{cav_review} reactive to membrane stress\cite{cav_stress, woodman}. Striking experiments, showing {\em Caveolae} flattening under tension\cite{flat_cav}, support the idea that the delivery of invaginated membrane area to the plasma membrane is controlled by tension\cite{cav_reservoir}. The shape of membrane domains in artificial membrane systems, such as giant vesicles, is also known to be dependent on the membrane tension\cite{membr_domains1}. Indeed, domain budding has been observed upon the decrease of membrane stress\cite{membr_domains2}.

\begin{figure}[h]
\centerline{\includegraphics[width=8cm]{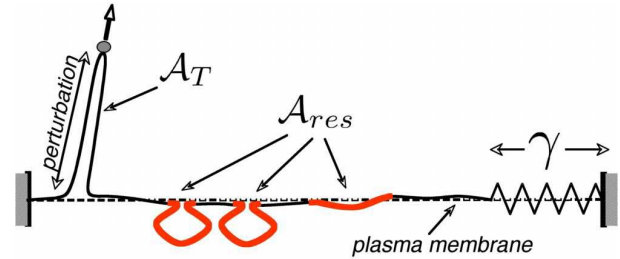}}
\caption{\label{modelfig} \small  Sketch depicting the mechanical response of a membrane to the pulling of a membrane tether. Pulling an area $\A_T$ out of the cell either increases the membrane tension $\gamma$ (the length of the spring), or triggers membrane exchange with a reservoir (flattening of invaginated membrane domains)}
\end{figure}

Our model is summarised in \fig{modelfig}. Here the cytoskeleton is included as linear elastic element (spring) mechanically coupled to the cell membrane. Although crude, this approximation is consistent with the experimental results of \cite{sheetz_buffer} (see \cite{SUPP}). Furthermore, it allows us isolate and study the response of the reservoir, which can be thought of as one element in a chain of response that transmits stress to the cytoskeleton. Cytoskeleton elasticity and membrane cytoskeleton anchoring are reflected by an effective stretching coefficient $K_s$ of order $5.10^4k_BT/\mu  m^4$\cite{SUPP}. Within this framework, the variation of the cell tension with the tether area $\A_T$ reads:
\be
\gamma=\gamma_0+K_s\left(\A_T+\A_{res}-\A_{res}^{(0)}\right)
\label{lineartension}
\ee
where $\gamma_0$ and $\A_{res}^{(0)}$ are the cell tension and reservoir area at rest (without tether, $\A_T=0)$. If membrane area is delivered to the plasma membrane, as during exocytosis, the ``tether'' area is negative. One can see that the membrane tension $\gamma$ can be maintained  constant upon tether pulling only if the decrease of reservoir area matches the increase of tether area. 

\begin{figure}[h]
%\vspace{0.5cm}
\centerline {\includegraphics[width=7cm]{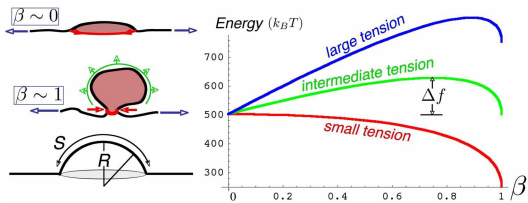}}
\caption{\label{domainfig} \small {\bf Left}:  A sketch of a flat and an invaginated domain, above our idealization for the domain shape; a spherical cap of area $S$, radius $R$ (curvature $C=2/R$). The shape is defined by a single parameter $0<\beta= SC^2/16\pi<1$. {\bf Right}: energy of a domain for increasing value of the membrane tension. Low tensions (red) favor budded shape ($\beta=1$), and high tensions (blue) favor flat shapes ($\beta=0$). At coexistence (green), flat and budded domains have the same energy, and the tension is given by \eq{regul0} The budded state remains a local minimum for high membrane tension  and there exists an energy barrier $\Delta f$ to domain flattening.}
%\vspace{0.5cm}
\end{figure}

The membrane domains constituting the reservoir are described within a very general theoretical framework. It is based on the domain bending rigidity, that disfavors the budded state, and its composition difference with the rest of the membrane, that promotes domain invagination in order to reduce the length of the domain periphery (\fig{domainfig}).  Domains are treated as spherical caps of fixed area $S$ and adjustable curvature $C$. Their shape is uniquely characterized by the parameter $\beta$ (\fig{domainfig} and \eq{fsk}), equal to unity for a fully budded domain (a sphere), and that vanishes for a flat domain. The energy $f$ of a domain contains a surface tension-independent part $\fsk$. This itself involves a term arising from the line tension $\sigma$  \cite{membr_domains1}, proportional to the length of the cap edge (neck), and a term giving the bending energy of the cap, proportional to the bending rigidity $\kappa$ and the squared curvature\cite{safran}. Simple geometry then gives
\be
\fsk[\beta]=\sqrt{\pi S}\sigma\sqrt{1-\beta}+8\pi\kappa\beta\qquad\beta=SC^2/16\pi
\label{fsk}
\ee
If the domain size exceeds a critical value $S_c=\pi(4\kappa/\sigma)^2$, an invaginated sphere  ($\beta=1$) has a lower energy than  a flat domain ($\beta=0$) and budding is expected\cite{lipowsky_budding}. We assume that $S>S_c$ in what follows. Typical values of the parameters $\sigma\sim k_BT/nm$ and $\kappa=20k_BT$ correspond to a critical size $S_c=(100nm)^2$, similar to the size of caveolae. In practice invaginated domains remain attached to the mother membrane by a small neck. Here, this is controlled phenomenologically  by assigning the value $\ab$ ($\gg1$) to the ratio of invagination radius $R$ to neck size, giving an upper bound $\bb=1-(\ab/2)^2$ to the shape parameter. 

Including the membrane tension $\gamma$, the domain energy reads: $f[\beta]=\fsk[\beta]+\gamma S\beta$. Increasing membrane tension increases the energy of curved states $\beta>0$ and promotes the flat state (see \fig{domainfig}). The flat states eventually becomes stable for a critical tension $\gamma^*$:
\be
S\gamma^*=-(\fsk[1]-\fsk[0])=2\sqrt{\pi S}\sigma-8\pi\kappa
\label{regul0}
\ee
As can be seen in \fig{domainfig}, the budded and flat domain shapes are separated  by an energy barrier for intermediate tension.  The existence of this barrier is crucial to their function as tension regulators, as it allows the coexistence of flat and invaginated domains. Variation of the membrane strain (the tether length) may occur without change of tension by adjusting the fraction of budded domains.

The energy scales in this system (with $S=0.1\mu m^2$) are
\be
\bsigma\equiv\sqrt{\pi S}\sigma\simeq\bkappa\equiv8\pi\kappa\simeq 500k_BT\ ;\ \bgamma\equiv\gamma S
\label{scales}
\ee 
The energy of surface tension competes with the line and bending energies for $\bgamma\simeq 500k_BT$, or $\gamma=2\ 10^{-5}J/m^2$ (corresponding to tether forces of order $10pN$). This is precisely in the range of mechanical tension recorded for cellular membranes\cite{sheetz_cytoskel}, which is very encouraging for the biological relevance of our model. One notes that the energy scale is very large compared to the thermal energy $k_BT$, or to the energy of any ``active temperature'' present in biological systems\cite{membrane_activity}. This has two important physical consequences: {\em (i)} the shape transition of a domain is very discontinuous, a domain snaps open rather than continuously flattening upon tension increase, and {\em (ii)}  the budding and flattening transitions should actually occur at different tensions, for which the respective energy barriers are of order $k_BT$. In biological systems, the ``temperature'' $T$ might be seen as a parameter reflecting cellular activity, such as the polymerization of the actin cortex near the membrane, and the activity of membrane pumps. For simple cells such as Red Blood Cells, it is typically a few times the thermodynamic temperature\cite{membrane_activity}.

The bottleneck for the shape transition is the maximum of energy, which corresponds to a shape parameter $\beta_{max}=1-(\bsigma/(\bkappa+\bgamma))^2$. The budding  and flattening tensions ($\gamma^{(1)}$ and $\gamma^{(0)}$ respectively) at which the corresponding energy barrier vanishes are:
\be
\bgamma^{(1)}=\bsigma-\bkappa<\bgamma^*=2\bsigma-\bkappa<\bgamma^{(0)}=2\ab\bsigma-\bkappa
\label{trans}
\ee
where $\ab$ characterize the size of the invagination neck (see above). 

We investigate the tension regulation performed by a collection of $\N$  domains, of total area $\N S\bb$ (where $\bb=1-(\ab/2)^2$, is the largest value of shape parameter consistent with the existence of a finite size neck). 

When flat and budded domains coexist, a fraction $\epsilon$ of domains are invaginated, and the reservoir area  is $\A_{res}=S\N\epsilon\bb$. The total membrane energy, including the contribution $\fsk$ of each of the $\N$ membrane domains (\eq{fsk}) and the total work done against membrane tension can be written:
\be
\F=\N\left(\epsilon\fsk[\bb]+(1-\epsilon)\fsk[0]\right)+\int d\A\gamma[\A]
\label{ftot0}
\ee

Optimizing the energy for the fraction of invaginated domains $\partial\F/\partial\epsilon=0$ leads directly to the regulation of membrane tension, when flat and budded domains coexist ($0<\epsilon<1$). Substituting in \eq{ftot0} the expression for the surface tension \eq{lineartension} (with $\bb\lesssim1$), we find that the tension is set to the value $\gamma^*$ of \eq{regul0}, which depends on the characteristics of the membrane reservoir ($\sigma$, $\kappa$, and $S$), but {\em not}  on the tether area. Regulation is achieved by adjusting the fraction of budded domains to:
\be
\epsilon^*=\epsilon_0-\frac{(\bgamma_0+\bks\A_T/S)-\bgamma^*}{\bks\N}\qquad\bks\equiv K_s S^2
\label{regulGE0}
\ee
where $\epsilon_0$ is the fraction of budded domains corresponding to the tension at rest $\gamma_0$, and where a normalized stretching coefficient $\bks(\sim0.1\bsigma)$, with dimension of energy, is introduced for convenience. 

If the reservoir is given time to equilibrate,  regulation starts for a level of perturbation corresponding to a tether area $\A_T^{(1)}$ (at which all domains are budded, $\epsilon^*=1$), and ends at a tether area $\A_T^{(0)}$ (at which all domains are flat, $\epsilon^*=0$).  The tension of the cell membrane is then set to the values $\gamma^*$, for any perturbation within the range $\A_T^{(1)}<\A_T<\A_T^{(0)}$. If the perturbation is very fast, one expect a large difference between the regulated tension upon tether pulling and tether retraction, in agreement with \eq{trans}.

To obtain the full kinetic response of the membrane to strain, we describe the transition as a classical Kramers' process\cite{vankampen}, were the transition time between two states is exponential with the energy barrier $\Delta f$ that has to be overcome in the process: $\tau=\tau_0 \exp[\Delta f/k_BT]$. Here, $\tau_0$ is the characteristic fluctuation time of the domain shape, and $T$ may be an effective temperature resulting from the large biochemical activity near the cell membrane\cite{membrane_activity}. The transition time is very much dependent upon the membrane tension.  Assuming that the transition of a single domain occurs with negligible  change of tension (this implies $\N\gg1$), the transition is fully described by the energy $f[\beta]=\fsk[\beta]+\bgamma\beta$, with $\gamma$ given by \eq{lineartension}. Assuming for simplicity that both domain flattening and budding involve the same fluctuation time $\tau_0$, the kinetic evolution of the fraction $\epsilon$ is given by
\be
\tau_0\frac{d\epsilon}{dt}=-\epsilon e^{-\frac{f_{max}-f[\bb]}{k_BT}}+(1-\epsilon) e^{-\frac{f_{max}-f[0]}{k_BT}}
\label{kineq}
\ee
where the maximum of energy $f_{max}$ corresponds to the least favorable domain shape $\beta_{max}=1-\bsigma^2/(\bkappa+\bgamma)^2$

In order to mimic the tether pulling experiment, where the tether is typically extracted at constant speed ($4\mu m/s$) in \cite{sheetz_buffer}, we consider the reservoir response to a perturbation applied with a given rate $\dot\A_T$: $\A_T=\A_T^{(0)}+\dot\A_T t$. If the perturbation is applied slowly ($\dot\A_T\tau_0\ll\A_{res}$), the reservoir has time to equilibrate ($d\epsilon/dt=0$), and the fraction $\epsilon^*_T$ of budded domain is found from \eq{kineq} to be given by $\epsilon^*_T/(1-\epsilon^*_T)=e^{(f[0]-f_{bud})/(k_BT)}$, with$f[0]-f_{bud}\simeq2\bsigma-(\bgamma(\epsilon)+\bkappa)$. 
The fraction  $\epsilon_T$ is the equivalent of the equilibrium fraction $\epsilon^*$ (\eq{regulGE0}), that takes thermal fluctuations into account ($\epsilon_T^*\simeq\epsilon^*$ if $\bsigma\gg k_BT$). Thermal fluctuations smoothen the transition between budded and flat domains by allowing states of non-minimal energy to be populated. As a consequence, the  tension is not perfectly constant during the transition, and the slope at mid plateau is of order  $\partial\gamma^*/\partial\A_T{}_{|_{plat}}\simeq 4k_BT/(S\A_{res})$. If, on the other hand, the perturbation is applied very fast, the shape transition requires small energy barriers, which means high tension for bud flattening, and low tension for domain budding, respectively $\gamma^{(0)}$ and $\gamma^{(1)}$ give by \eq{trans}.  

 \begin{figure}[h]
\centerline{\includegraphics[width=8cm]{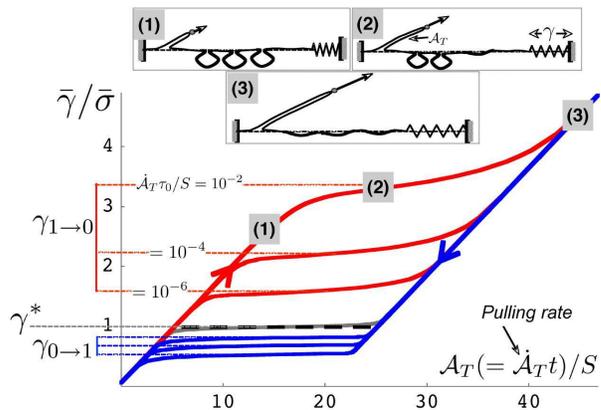}}
\caption{\label{kinfig} \small Sketch of the change in membrane morphology upon tether extraction. At low strain (1) all membrane domains are budded, and the membrane tension $\gamma$ increases linearly with the membrane area $\A_T$. For large perturbation (3) all domains are flat and a similar linear increase is observed. In between (2) the membrane tension is maintained at a plateau value while flat and budded domains coexist. The equilibrium reservoir response ($\gamma^*$, dashed gray line) corresponds to a quasi-static perturbation. The dynamical responses at constant rate $\dot\A_T$ is shown upon extraction (red) and retraction (blue) for various rates. The hysteresis, which increases with the perturbation rate, illustrates  the kinetic nature of the domain shape transformation}
\end{figure}

The physical mechanism at the origin of tension regulation and the membrane hysteretic response to tether extraction and retraction are shown in \fig{kinfig}. To obtain an analytical expression of the plateau height with the perturbation rate, we approximate that the tension is almost constant during regulation ($d\bgamma/dt\sim0$) so that  the energy barrier is of order $\Delta f/(k_BT)\sim\log{\A_{res}/(\dot \A_T\tau_0)}$. The plateau tension upon increase and decrease of the perturbation are then respectively given by $\bgamma_{1\rightarrow0}=\bgamma^{(0)}- 2\ab^{3/2}\sqrt{\bsigma k_BT \log}$ and $\bgamma_{0\rightarrow1}=\bgamma^{(1)}+\sqrt{\bsigma k_BT \log}$, with $\log\equiv \log[\A_{res}/(2\dot \A_T\tau_0)]$. As expected for activated processes, the dependence of the tension at transition with the rate of perturbation $\dot \A_T$ is logarithmic. The same is true for the slope of the plateau, which can be estimated by identifying the plateau inflexion point. The condition $d^2\bgamma/dt^2=0$ imposes $\dot{\Delta f}/(k_BT)\simeq\dot\epsilon/\epsilon$, corresponding to a plateau slope  $\partial\bgamma_{1\rightarrow0}/\partial\A_T{}_{|_{plat}}\simeq \sqrt{2k_BT\bsigma\ab^3/\log}/\A_{res}$. As the effective membrane temperature may be inferred independently from measurement of membrane fluctuations\cite{membrane_activity}, the study of both the height and slope of the tension plateau gives valuable information of the kinetics of area transfer between the reservoir and the plasma membrane. 

The difference between the quasi-static and dynamic plateau can be of order $10^{-4}J/m^2$ and correspond to a difference in force of order $10pN$. This is precisely the scale of the forces measured upon tether extraction in \cite{sheetz_buffer}. This possibility thus exists that the initial increase of tension observed prior to the plateau in \cite{sheetz_buffer} is of purely kinetic origin and originate from the slow response of an already partially unfolded reservoir at rest. In this case, the membrane tension of the cell at rest would be $\gamma^*$ of \eq{trans}, fully controlled by the mechanical properties of the membrane domains forming the reservoir\cite{SUPP}.

In summary, we have derived the mechanical reactivity of a cell membrane, in contact with a reservoir composed of invaginated membrane domains liable to flatten under strain. The flattening transition is first order, which means that the invaginations snap open above a critical strain rather than continuously flattening, an observation consistent with experimental evidences from the membrane invagination caveolae\cite{cav_reservoir}. As a consequence, the cell mechanical response shows a plateau during the transition, corresponding to the coexistence of flat and invaginated domains. This study provides the basis for a mechanical regulation of the tension of the cell membrane. The mechanism at the origin of this regulation is of  purely physical origin, consistent with the fast timescale ($< 0.1 sec$) of the observed cellular response. In practice, a ring of specialized membrane proteins such as dynamin\cite{cav-dynamin} is often present at the neck of membrane invagination. These proteins most probably influence the domain line energy, and might even dominate the energy required to flatten the domain. The regulation of membrane tension by membrane invaginations rely on the existence of two well defined domain shapes, separated by an energy barrier. If anything, neck proteins can only increase the energy barrier to flattening, thereby reinforcing  tension regulation.

 This work also opens the possibility of a new quantitative ``force spectroscopy'' of the cell membrane. One could thereby obtain  structural information on the membrane organization, in much the same way information on a protein structure can be gathered from force measurement upon protein unfolding\cite{protein_unfolding}. As a first step, one may identify the fairly regular oscillation of the force during regulation in  \cite{sheetz_buffer}, to the flattening of single domains. Preliminary analysis\cite{SUPP} hints at domains of area $S\sim (400nm)^2$.


\begin{thebibliography}{xx}


\bibitem{sheetz}  D. Raucher, \& M.P. Sheetz   (1999) {\it J. Cell Biol.} {\bf 144}, 497-506\qquad D. Raucher, \& M.P. Sheetz (2000) {\it J. Cell Biol.} {\bf 148}, 127-136

\bibitem{sheetz_cytoskel} M.P. Sheetz   (2001)  {\it Nature Rev. Mol. Cell Bio.} {\bf 2}, 392-396

\bibitem{sheetz_pip2} D. Raucher, T. Stauffer, W. Chen, K. Shen, S. Guo, JD. York, MP. Sheetz, and T. Meyer. (2000)
{\em Cell} {\bf 100}, 221Ð228

\bibitem{morris_SAR} C.E. Morris \& U. Homann  (2000) {\it J. Membrane Biol.} {\bf 179}, 79-102

\bibitem{sheetz_buffer} D. Raucher, \& M.P. Sheetz  (1999)  {\it Biophys. J.} {\bf 77}, 1992-2002

\bibitem{cav_review} K. Simons \& E. Ikonen (1997) {\it Nature} 387, 569-572  \quad Schlegel, A., D. Volont\'e, J.A.  Engelman, F. Galbiati, P. Mehta, X.L. Zhang, P.E. Scherer, and M.P. Lisanti.  (1998) {\em Cell Signal} 10:457-463

\bibitem{cav_stress}  H. Park, Y.-M. Go, P.L. St.John, M.C. Maland, M.P. Lisanti, D.R. Abrahamson, \& H. Jo  (1998) {\it J. Biological Chemistry} {\bf 273} 32304-32311 

\bibitem{woodman} 	S.E. Woodman, F. Sotgia, F. Galbiati, C. Minetti \& Lisanti, M. P.  (2004){\it Neurology} {\bf 62} p.538-543. 

\bibitem{flat_cav} L. Prescott \& M.W. Brightman. (1976) {\it Tissue Cell.} {\bf 8} pp.248-58 

\bibitem{cav_reservoir} A.F. Dulhunty and C. Franzini-Armstrong (1975) {\em J Physiol.} {\bf 250}, 513-39.

\bibitem{membr_domains1} T. Baumgart, S. T. Hess, \& W. W. Webb (2003) {\it Nature} {\bf 425} p. 821-824\quad G. Staneva, M. I. Angelova, \& K. Koumanov  (2004)  {\it Chemistry and Physics of Lipids} {\bf 129} p. 53Ð62

\bibitem{membr_domains2}  A. Roux \& P. Bassereau {\em private communication}.

\bibitem{SUPP} P. Sens \& M.S. Turner, {\em in preparation}

\bibitem{safran} S.A. Safran 1994. Statistical Thermodynamics of Surfaces, Interfaces and  Membranes. Perseus, Cambridge, MA.

\bibitem{lipowsky_budding} R. Lipowsky (1992) {\it J.  Phys. II France} {\bf 2} 1825-1840  \quad R. Lipowsky  (1993) {\it Biophys. J.} {\bf 64} 1133-1138

\bibitem{membrane_activity}  S. Tuvia, A. Almagor, A. Bitler, S. Levin, R. Korenstein \& S. Yedgar   (1997) {\em Proc. Natl. Acad. Sci. USA} {\bf 94}, pp. 5045Ð5049\quad N.Gov, A. Zilman, \& S.A. Safran (2005) {\em Cond-Mat} 0207514

\bibitem{vankampen} N.G. Van Kampen (1992) {\em Stochastic processes in Physics and Chemistry}, Elsevier, Amsterdam

\bibitem{evans_binding} E. Evans (2001) {\em Annu. Rev. Biophys. Biomol. Struct.}{\bf 30} pp105-28

\bibitem{cav-dynamin} Q. Yao, J. Chen, H. Cao, J.D. Orth, J.M. McCaffery, R.V. Stan \& M.A. McNiven MA. (2005) {\em J Mol Biol.} {\bf 29}, 491-501. 

\bibitem{protein_unfolding} M. Rief, M. Gautel, F. Oesterhelt, J.M. Fernandez \& H.E. Gaub (1997) {\em Science} {\bf 276}, 1109Ð1112



\end{thebibliography}
\end{document}